\documentclass[12pt]{article}
\usepackage{graphicx}
\usepackage{epsfig}
\usepackage{amssymb}

\newcommand{\Li}{\mathop{\mathrm{Li}}\nolimits}

\begin{document}

\boldmath
\title{
\vskip-3cm{\baselineskip14pt
\centerline{\normalsize DESY 08-002\hfill ISSN 0418-9833}
\centerline{\normalsize January 2008\hfill}}
\vskip1.5cm
\bf Heavy-quark contributions to the ratio $F_L/F_2$ at low $x$}
\unboldmath

\author{
Alexey~Yu.~Illarionov\thanks{Electronic address:
{\tt illario@sissa.it}.}\\
{\normalsize\it Scuola Internazionale Superiore di Studi Avanzati,
Via Beirut, 2--4, 34014 Trieste, Italy}\\
\\
Bernd A. Kniehl\thanks{Electronic address:
{\tt kniehl@desy.de}.},
Anatoly~V.~Kotikov\thanks{Electronic address:
{\tt kotikov@theor.jinr.ru};
on leave of absence from the Bogoliubov Laboratory of
Theoretical Physics,
Joint Institute for Nuclear Research, 141980 Dubna, Moscow region, Russia.}\\
{\normalsize\it II. Institut f\"ur Theoretische Physik,
Universit\"at Hamburg,}
\\
{\normalsize\it Luruper Chaussee 149, 22761 Hamburg, Germany}}

\date{}

\maketitle

\begin{abstract}
We study the heavy-quark contribution to the proton structure functions
$F_2^i(x,Q^2)$ and $F_L^i(x,Q^2)$, with $i=c,b$, for small values of Bjorken's
$x$ variable at next-to-lading order and provide compact formulas for their
ratios $R_i=F_L^i/F_2^i$ that are useful to extract $F_2^i(x,Q^2)$ from
measurements of the doubly differential cross section of inclusive
deep-inelastic scattering at DESY HERA.
Our approach naturally explains why $R_i$ is approximately independent of $x$
and the details of the parton distributions in the small-$x$ regime.

\medskip

\end{abstract}

\newpage

\section{Introduction}
\label{sec:intro}

The totally inclusive cross section of deep-inelastic lepton-proton
scattering (DIS) depends on the square $s$ of the centre-of-mass energy,
Bjorken's variable $x=Q^2/(2pq)$, and the inelasticity variable $y=Q^2/(xs)$,
where $p$ and $q$ are the four-momenta of the proton and the virtual photon,
respectively, and $Q^2=-q^2>0$.
The doubly differential cross section is parameterized in terms of the
structure function $F_2$ and the longitudinal structure function $F_L$, as
\begin{equation}
\frac{d^2 \sigma}{dx\,dy}=\frac{2\pi\alpha^2}{xQ^4}
\{[1+(1-y)^2]F_2(x,Q^2)-y^2 F_L(x,Q^2)\},
\end{equation}
where $\alpha$ is Sommerfeld's fine-structure constant.
At small values of $x$, $F_L$ becomes non-negligible and its contribution
should be properly taken into account when the $F_2$ is extracted from the
measured cross section.
The same is true also for the contributions $F_2^i$ and $F_L^i$ of $F_2$ and
$F_L$ due to the heavy quarks $i=c,b$.

Recently, the H1 \cite{Adloff:1996xq,Aktas:2004az,Aktas:2005iw} and ZEUS 
\cite{Breitweg:1997mj,Chekanov:2003rb,Chekanov:2007ch}
Collaborations at HERA presented new data on $F_2^c$ and $F_2^b$.
At small $x$ values, of order $10^{-4}$, $F_2^c$ was found to be around
$25\%$ of $F_2$, which is considerably larger than what was observed by the
European Muon Collaboration (EMC) at CERN \cite{Aubert:1982tt} at larger $x$
values, where it was only around $1\%$ of $F_2$.
Extensive theoretical analyses in recent years have generally served to
establish that the $F_2^c$ data can be described through the perturbative
generation of charm within QCD (see, for example, the review in
Ref.~\cite{Cooper-Sarkar:1997jk} and references cited therein).

In the framework of Dokshitzer-Gribov-Lipatov-Altarelli-Parisi (DGLAP)
dynamics \cite{Gribov:1972ri}, there are two basic methods to study
heavy-flavour physics.
One of them \cite{Kniehl:1996we} is based on the massless evolution of parton
distributions and the other one on the photon-gluon fusion (PGF) process
\cite{Lopez:1979bb}.
There are also some interpolating schemes (see
Ref.~\cite{Olness:1987ep} and references cited therein).
The present HERA data on $F_2^c$
\cite{Adloff:1996xq,Aktas:2004az,Aktas:2005iw,Breitweg:1997mj,Chekanov:2003rb,Chekanov:2007ch}
are in good agreement with the modern theoretical predictions.

In earlier HERA analyses \cite{Adloff:1996xq,Breitweg:1997mj}, $F_L^c$ and
$F_L^b$ were taken to be zero for simplicity.
Four years ago, the situation changed:
in the ZEUS paper \cite{Chekanov:2003rb}, the $F_L^c$ contribution at
next-to-leading order (NLO) was subtracted from the data; 
in Refs.~\cite{Aktas:2004az,Aktas:2005iw}, the H1 Collaboration introduced
the reduced cross sections 
\begin{equation}
\tilde{\sigma}^{i\overline{i}}=\frac{xQ^4}{2\pi\alpha^2[1+(1-y)^2]}\,
\frac{d^2\sigma^{i\overline{i}}}{dx\,dy}
=F_2^i(x,Q^2)-\frac{y^2}{1+(1-y)^2}F_L^i(x,Q^2)
\label{eq:red}
\end{equation}
for $i=c,b$ and thus extracted $F_2^i$ at NLO by fitting their data.
Very recently, a similar analysis, but for the doubly differential 
cross section $d^2 \sigma^{i\overline{i}}/(dx\,dy)$ itself, has been performed
by the ZEUS Collaboration \cite{Chekanov:2007ch}.

In this letter, we present a compact formula for the ratio $R_i=F_L^i/F_2^i$,
which greatly simplifies the extraction of $F_2^i$ from measurements of
$d^2 \sigma^{i\overline{i}}/(dx\,dy)$.

\section{Master formula}
\label{sec:approach}

We now derive our master formula for $R_i(x,Q^2)$ appropriate for small values
of $x$, which has the advantage of being independent of the parton
distribution functions (PDFs) $f_a(x,Q^2)$, with parton label
$a=g,q,\overline{q}$, where $q$ generically denotes the light-quark flavours.
In the small-$x$ range, where only the gluon and quark-singlet contributions
matter, while the non-singlet contributions are negligibly small, we
have\footnote{%
Here and in the following, we suppress the variables $\mu$ and $m_i$ in the
argument lists of the structure and coefficient functions for the ease of
notation.}
\begin{equation}
F_k^i(x,Q^2)=\sum_{a=g,q,\overline{q}}\sum_{l=+,-}
C_{k,a}^l(x,Q^2)\otimes xf_a^l(x,Q^2),
\label{eq:pm}
\end{equation}
where $l=\pm$ labels the usual $+$ and $-$ linear combinations of the gluon
and quark-singlet contributions, $C_{k,a}^l(x,Q^2)$ are the DIS coefficient
functions, which can be calculated perturbatively in the parton model of QCD,
$\mu$ is the renormalization scale appearing in the strong-coupling constant
$\alpha_s(\mu)$, and the symbol $\otimes$ denotes convolution according to
the usual prescription, $f(x)\otimes g(x)=\int_x^1(dy/y)f(y)g(x/y)$.
Massive kinematics requires that $C_{k,a}^l=0$ for $x>b_i=1/(1+4a_i)$, where
$a_i=m_i^2/Q^2$.
We take $m_i$ to be the solution of $\overline{m}_i(m_i)=m_i$, where
$\overline{m}_i(\mu)$ is defined in the modified minimal-subtraction
($\overline{\mathrm{MS}}$) scheme.

Exploiting the small-$x$ asymptotic behaviour of $f_a^l(x,Q^2)$
\cite{Lopez:1979bb},
\begin{equation}
f_a^l(x,Q^2)\stackrel{x\to0}{\to}\frac{1}{x^{1+\delta_l}},
\end{equation}
Eq.~(\ref{eq:pm}) can be rewritten as
\begin{equation}
F_k^i(x,Q^2)\approx\sum_{a=g,q,\overline{q}}\sum_{l=+,-}
M_{k,a}^l(1+\delta_l,Q^2)xf_a^l(x,Q^2),
\label{eq:pm1}
\end{equation}
where
\begin{equation}
M_{k,a}^l(n,Q^2)=\int_0^{b_i}dx\,x^{n-2}C_{k,a}^l(x,Q^2)
\label{eq:mel}
\end{equation}
is the Mellin transform, which is to be analytically continued from integer
values $n$ to real values $1+\delta_l$.

As demonstrated in Ref.~\cite{Kotikov:1998qt}, HERA data support the modified
Bessel-like behavior of PDFs at low $x$ values predicted in the framework of
the so-called generalized double-asymptotic scaling regime.
In this approach, one has $M_{k,a}^+(1,Q^2)=M_{k,a}^-(1,Q^2)$ if 
$M_{k,a}^l(n,Q^2)$ are devoid of singularities in the limit $\delta_l\to0$, as
in our case.
Defining $M_{k,a}(1,Q^2)=M_{k,a}^\pm(1,Q^2)$ and using
$f_a(x,Q^2)=\sum_{l=\pm}f_a^l(x,Q^2)$, Eq.~(\ref{eq:pm1}) may be simplified to
become
\begin{equation}
F_k^i(x,Q^2)\approx\sum_{a=g,q,\overline{q}}M_{k,a}(1,Q^2)xf_a(x,Q^2).
\label{eq:pm2}
\end{equation}
A further simplification is obtained by neglecting the contributions due to
incoming light quarks and antiquarks in Eq.~(\ref{eq:pm2}), which is justified
because they vanish at LO and are numerically suppressed at NLO for small
values of $x$.
One is thus left with the contribution due to PGF \cite{Lopez:1979bb},
\begin{equation}
F_k^i(x,Q^2)\approx M_{k,g}(1,Q^2)xf_g(x,Q^2).
\label{eq:pm3}
\end{equation}
In fact, the non-perturbative input $f_g(x,Q^2)$ does cancels in the ratio
\begin{equation}
R_i(x,Q^2)\approx\frac{M_{L,g}(1,Q^2)}{M_{2,g}(1,Q^2)},
\label{eq:ri}
\end{equation}
which is very useful for practical applications.
Through NLO, $M_{k,g}(1,Q^2)$ exhibits the structure
\begin{eqnarray}
M_{k,g}(1,Q^2)&=&e_i^2a(\mu)\left\{M_{k,g}^{(0)}(1,a_i)
+a(\mu)\left[M_{k,g}^{(1)}(1,a_i)+M_{k,g}^{(2)}(1,a_i)
\vphantom{\frac{\mu^2}{m_i^2}}\right.\right.
\nonumber\\
&&{}\times\left.\left.\ln\frac{\mu^2}{m_i^2}
\right]\right\}+{\mathcal O}(a^3),
\label{eq:exp}
\end{eqnarray}
where $e_i$ is the fractional electric charge of heavy quark $i$ and
$a(\mu)=\alpha_s(\mu)/(4\pi)$ is the couplant.
Inserting Eq.~(\ref{eq:exp}) into Eq.~(\ref{eq:ri}), we arrive at our master
formula
\begin{eqnarray}
R_i(x,Q^2)&\approx&\frac{M_{L,g}^{(0)}(1,a_i)+a(\mu)
\left[M_{L,g}^{(1)}(1,a_i)+M_{L,g}^{(2)}(1,a_i)\ln(\mu^2/m_i^2)\right]}
{M_{2,g}^{(0)}(1,a_i)+a(\mu)
\left[M_{2,g}^{(1)}(1,a_i)+M_{2,g}^{(2)}(1,a_i)\ln(\mu^2/m_i^2)\right]}
\nonumber\\
&&{}+{\mathcal O}(a^2).
\label{eq:master}
\end{eqnarray}
We observe that the right-hand side of Eq.~(\ref{eq:master}) is independent of
$x$, a remarkable feature that is automatically exposed by our procedure.
In the next two sections, we present compact analytic results for the LO
($j=0$) and NLO ($j=1,2$) coefficients $M_{k,g}^{(j)}(1,a_i)$, respectively.

\section{LO results}
\label{sec:lo}

The LO coefficient functions of PGF can be obtained from the QED case
\cite{Baier:1966bf} by adjusting coupling constants and colour factors, and
they read \cite{Witten:1975bh,Kotikov:2001ct}
\begin{eqnarray}
C_{2,g}^{(0)}(x,a) &=& -2x\{[1-4x(2-a)(1-x)]\beta
-[1-2x(1-2a)
\nonumber\\
&&{}+2x^2(1-6a-4a^2)]L(\beta)\},
\nonumber \\
C_{L,g}^{(0)}(x,a) &=&  8 x^2[(1-x)\beta-2ax L(\beta)],
\end{eqnarray}
where
\begin{equation}
\beta=\sqrt{1-\frac{4ax}{1-x}},\qquad
L(\beta)=\ln\frac{1+\beta}{1-\beta}.
\end{equation}
Using the auxiliary formulas
\begin{eqnarray}
\int_0^b x^m \beta&=& 
\left\{\begin{array}{ll}
1-2aJ(a), & \mbox{if}~m=0 \\
\frac{b}{2}[1-2a-4a(1+3a)J(a)], & \mbox{if}~m=1 \\
\frac{b^2}{3}[(1+3a)(1+10a)-6a(1+6a+10a^2)J(a)], & \mbox{if}~m=2 \\
\end{array}\right.,\qquad
\label{eq:bet}\\
\int_0^b x^m L(\beta) &=& 
\left\{\begin{array}{ll}
J(a), & \mbox{if}~m=0 \\
-\frac{b}{2}[1-(1+2a)J(a)], & \mbox{if}~m=1 \\
-\frac{b^2}{3}[3(1+2a)-2(1+4a+6a^2)J(a)], & \mbox{if}~m=2
\end{array}\right.,
\end{eqnarray}
where
\begin{equation}
J(a) = - \sqrt{b}\ln t,\qquad t=\frac{1-\sqrt{b}}{1+\sqrt{b}},
\label{eq:ja}
\end{equation}
we perform the Mellin transformation in Eq.~(\ref{eq:mel}) to find
\begin{eqnarray}
M_{2,g}^{(0)}(1,a) &=& \frac{2}{3}[1+2(1-a)J(a)],
\nonumber\\
M_{L,g}^{(0)}(1,a) &=& \frac{4}{3}b[1+6a-4a(1+3a)J(a)].
\end{eqnarray}
At LO, the small-$x$ approximation formula thus reads
\begin{equation}
R_i\approx
2b_i\frac{1+6a_i-4a_i(1+3a_i)J(a_i)}{1+2(1-a_i)J(a_i)}.
\label{eq:lo}
\end{equation}

\section{NLO results}
\label{sec:nlo}

The NLO coefficient functions of PGF are rather lengthy and not published in
print; they are only available as computer codes \cite{Laenen:1992zk}.
For the purpose of this letter, it is sufficient to work in the high-energy
regime, defined by $a_i\ll1$, where they assume the compact form
\cite{Catani:1992zc}
\begin{equation}
C_{k,g}^{(j)}(x,a)=\beta R_{k,g}^{(j)}(1,a),
\label{eq:nlo}
\end{equation}
with
\begin{eqnarray}
R_{2,g}^{(1)}(1,a)&=&\frac{8}{9}C_A[5+(13-10a)J(a)+6(1-a)I(a)],
\nonumber\\
R_{L,g}^{(1)}(1,a)&=&-\frac{16}{9}C_A b
\{1-12a-[3+4a(1-6a)]J(a)+12a(1+3a)I(a)\},
\nonumber\\
R_{k,g}^{(2)}(1,a)&=&-4 C_A M_{k,g}^{(0)}(1,a),
\end{eqnarray}
where $C_A=N$ for the colour gauge group SU(N), $J(a)$ is defined by
Eq.~(\ref{eq:ja}), and 
\begin{equation}
I(a)=-\sqrt{b}\left[\zeta(2)+\frac{1}{2}\ln^2t-\ln(ab)\ln t+2\Li_2(-t)\right].
\end{equation}
Here, $\zeta(2)=\pi^2/6$ and
$\Li_2(x)=-\int_0^1(dy/y)\ln(1-xy)$ is the dilogarithmic function.
Using Eq.~(\ref{eq:bet}) for $m=0$, we find the Mellin 
transform~(\ref{eq:mel}) of Eq.~(\ref{eq:nlo}) to be
\begin{equation}
M_{k,g}^{(j)}(1,a)=[1-2aJ(a)]R_{k,g}^{(j)}(1,a).
\end{equation}
\section{Results}
\label{sec:results}

We are now in a position to explore the phenomenological implications of our
results.
As for our input parameters, we choose $m_c=1.25$~GeV and $m_b=4.2$~GeV.
While the LO result for $R_i$ in Eq.~(\ref{eq:lo}) is independent of the
unphysical mass scale $\mu$, the NLO formula~(\ref{eq:master}) does depend on
it, due to an incomplete compensation of the $\mu$ dependence of $a(\mu)$ by
the terms proportional to $\ln(\mu^2/Q^2)$, the residual $\mu$ dependence
being formally beyond NLO.
In order to estimate the theoretical uncertainty resulting from this, we put
$\mu^2=\xi Q^2$ and vary $\xi$.
Besides our default choice $\xi=1$, we also consider the extreme choice
$\xi=100$, which is motivated by the observation that NLO corrections are
usually large and negative at small $x$ values \cite{Salam:1998tj}.
A large $\xi$ value is also advocated in Ref.~\cite{Dokshitzer:1993pf}, where
the choice $\xi=1/x^a$, with $0.5<a<1$, is proposed.

\begin{table}
\caption{\label{tab:c}Values of $F_2^c(x,Q^2)$ extracted from the H1
measurements of $\tilde{\sigma}^{c\overline{c}}$ at low \cite{Aktas:2005iw}
and high \cite{Aktas:2004az} values of $Q^2$ (in GeV$^2$) at various values of
$x$ (in units of $10^{-3}$) using our approach at NLO for $\mu^2=\xi Q^2$ with
$\xi=1,100$.
The LO results agree with the NLO results for $\xi=1$ within the accuracy of
this table.
For comparison, also the results determined in
Refs.~\cite{Aktas:2004az,Aktas:2005iw} are quoted.}
\begin{tabular}{|cc|ccc|}
\hline
$Q^2$ & $x$ & H1 & $\mu^2=Q^2$ & $\mu^2=100\,Q^2$ \\
\hline
12 & 0.197 & $0.435\pm0.078$ & 0.433 & 0.432 \\
12 & 0.800 & $0.186\pm0.024$ & 0.185 & 0.185 \\
25 & 0.500 & $0.331\pm0.043$ & 0.329 & 0.329 \\
25 & 2.000 & $0.212\pm0.021$ & 0.212 & 0.212 \\
60 & 2.000 & $0.369\pm0.040$ & 0.368 & 0.368 \\
60 & 5.000 & $0.201\pm0.024$ & 0.200 & 0.200 \\
200 & 0.500 & $0.202\pm0.046$ & 0.201 & 0.201 \\
200 & 1.300 & $0.131\pm0.032$ & 0.130 & 0.130 \\
650 & 1.300 & $0.213\pm0.057$ & 0.212 & 0.213 \\
650 & 3.200 & $0.092\pm0.028$ & 0.091 & 0.091 \\
\hline
\end{tabular}
\end{table}
We now extract $F_2^i(x,Q^2)$ ($i=c,b$) from the H1 measurements of the
reduced cross sections in Eq.~(\ref{eq:red}) at low ($12<Q^2<60$~GeV$^2$)
\cite{Aktas:2005iw} and high ($Q^2>150$~GeV$^2$) \cite{Aktas:2004az} values of
$Q^2$ using the LO and NLO results for $R_i$ derived in Sections~\ref{sec:lo}
and \ref{sec:nlo}, respectively.
Our NLO results for $\mu^2=\xi Q^2$ with $\xi=1,100$ are presented for
$i=c,b$ in Tables~\ref{tab:c} and \ref{tab:b}, respectively, where they are
compared with the values determined by H1.
We refrain from showing our results for other popular choices, such as
$\mu^2=4m_i^2,Q^2+4m_i^2$ because they are very similar.
We observe that the theoretical uncertainty related to the freedom in the
choice of $\mu$ is negligibly small and find good agreement with the results
obtained by the H1 Collaboration using a more accurate, but rather cumbersome
procedure \cite{Aktas:2004az,Aktas:2005iw}.
The experimental data from the ZEUS Collaboration \cite{Chekanov:2007ch} do
not allow for such an analysis because they do not come in the form of
Eq.~(\ref{eq:red}).
\begin{table}
\caption{\label{tab:b}Values of $F_2^b(x,Q^2)$ extracted from the H1
measurements of $\tilde{\sigma}^{b\overline{b}}$ at low \cite{Aktas:2005iw}
and high \cite{Aktas:2004az} values of $Q^2$ (in GeV$^2$) at various values of
$x$ (in units of $10^{-3}$) using our approach at NLO for $\mu^2=\xi Q^2$ with
$\xi=1,100$.
The LO results agree with the NLO results for $\xi=1$ within the accuracy of
this table.
For comparison, also the results determined in
Refs.~\cite{Aktas:2004az,Aktas:2005iw} are quoted.}
\begin{tabular}{|cc|ccc|}
\hline
$Q^2$ & $x$ & H1 & $\mu^2=Q^2$ & $\mu^2=100\,Q^2$ \\
\hline
12 & 0.197 & $0.0045\pm0.0027$ & 0.0047 & 0.0046 \\
12 & 0.800 & $0.0048\pm0.0022$ & 0.0048 & 0.0048 \\
25 & 0.500 & $0.0123\pm0.0038$ & 0.0124 & 0.0124 \\
25 & 2.000 & $0.0061\pm0.0024$ & 0.0061 & 0.0061 \\
60 & 2.000 & $0.0190\pm0.0055$ & 0.0190 & 0.0190 \\
60 & 5.000 & $0.0130\pm0.0047$ & 0.0130 & 0.0130 \\
200 & 0.500 & $0.0413\pm0.0128$ & 0.0400 & 0.0400 \\
200 & 1.300 & $0.0214\pm0.0079$ & 0.0212 & 0.0212 \\
650 & 1.300 & $0.0243\pm0.0124$ & 0.0238 & 0.0238 \\
650 & 3.200 & $0.0125\pm0.0055$ & 0.0125 & 0.0125 \\
\hline
\end{tabular}
\end{table}

\begin{figure}
\begin{center}
\epsfig{figure=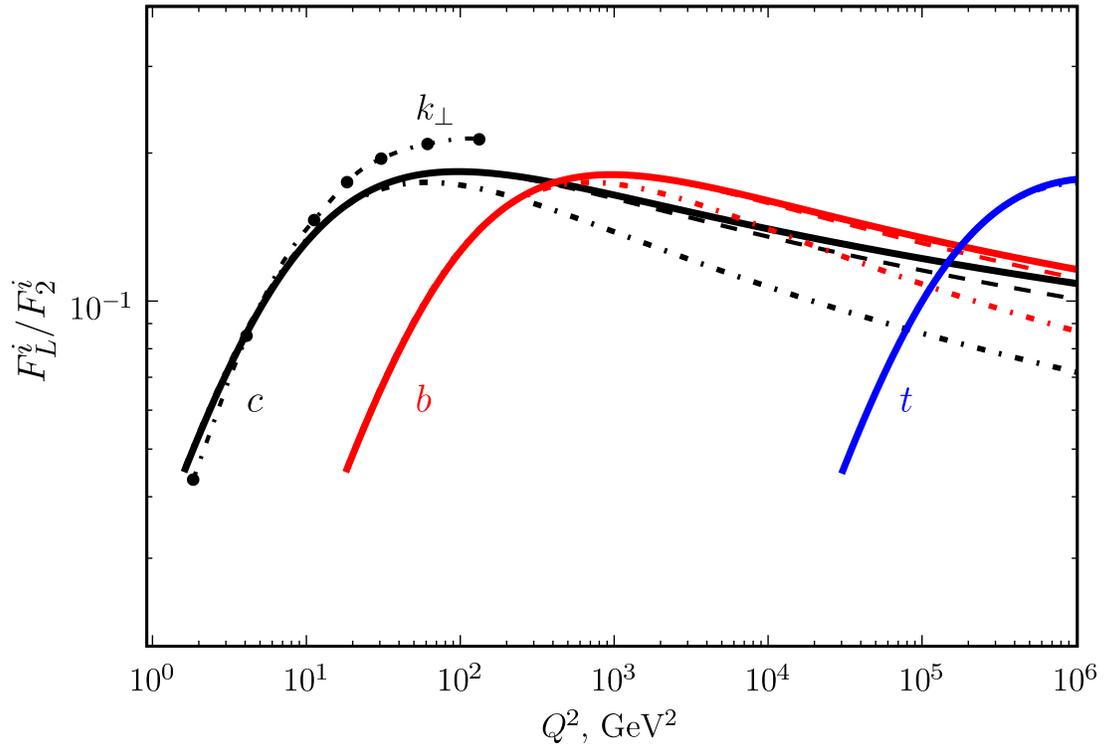,width=\linewidth}
\end{center}
\caption{$R_c$, $R_b$, and $R_t$ evaluated as functions of $Q^2$ at LO from
Eq.~(\ref{eq:lo}) (dot-dashed lines) and at NLO from Eq.~(\ref{eq:master})
with $\mu^2=4m_i^2$ (dashed lines) and $\mu^2=Q^2+4m_i^2$ (solid lines).
For comparison, the prediction for $R_c$ in the $k_t$-factorization approach
(dot-dot-dashed line) \cite{Kotikov:2001ct} is also shown.}
\label{fig:r}
\end{figure}

In order to assess the significance of and the theoretical uncertainty in the
NLO corrections to $R_i$, we show in Fig.~\ref{fig:r} the $Q^2$
dependences of $R_c$, $R_b$, and $R_t$ evaluated at LO from Eq.~(\ref{eq:lo})
and at NLO from Eq.~(\ref{eq:master}) with $\mu^2=4m_i^2,Q^2+4m_i^2$.
We observe from  Fig.~\ref{fig:r} that the NLO predictions are rather stable
under scale variations and practically coincide with the LO ones in the lower
$Q^2$ regime.
On the other hand, for $Q^2\gg4m_i^2$, the NLO predictions overshoot the LO
ones and exhibit an appreciable scale dependence.
We encounter the notion that the fixed-flavour-number scheme used here for
convenience is bound to break down in the large-$Q^2$ regime due to unresummed
large logarithms of the form $\ln(Q^2/m_i^2)$.
In our case, such logarithms do appear linearly at LO and quadratically at
NLO.
In the standard massless factorization, such terms are responsible for the
$Q^2$ evolution of the PDFs and do not contribute to the coefficient
functions.
In fact, in the variable-flavour-number scheme, they are 
$\overline{\mathrm{MS}}$-subtracted from the coefficient functions and
absorbed into the $Q^2$ evolution of the PDFs.
Thereafter, the asymptotic large-$Q^2$ dependences of $R_i$ at NLO should be
proportional to $\alpha_s(Q^2)$ and thus decreasing.
This is familiar from the Callan-Gross ratio $R=F_L/(F_2-F_L)$, as may be seen
from its $(x,Q^2)$ parameterizations in Ref.~\cite{GonzalezArroyo:1980wf}.
Fortunately, this large-$Q^2$ problem does not affect our results in
Tables~\ref{tab:c} and \ref{tab:b} because the bulk of the H1 data is located
in the range of moderate $Q^2$ values.
Furthermore, $R_i$ enters Eq.~(\ref{eq:red}) with the suppression factor
$y^2/[1+(1-y)^2]$.

The ratio $R_c$ was previously studied in the framework of the
$k_t$-factorization approach \cite{Kotikov:2001ct} and found to weakly depend
on the choice of unintegrated gluon PDF and to be approximately $x$
independent in the small-$x$ regime (see Fig.~8 in
Ref.~\cite{Kotikov:2001ct}). 
Both features are inherent in our approach, as may be seen at one glance from
Eq.~(\ref{eq:master}).
The prediction for $R_c$ from Ref.~\cite{Kotikov:2001ct}, which is included in
Fig.~\ref{fig:r} for comparison, agrees well with our results in the lower
$Q^2$ range, but it continues to rise with $Q^2$, while our results reach
maxima, beyond which they fall.
In fact, the $k_t$-factorization approach is likely to overestimate $R_c$ for
$Q^2\gg4m_i^2$, due to the unresummed large logarithms of the form
$\ln(Q^2/m_i^2)$ discussed above.

\section{Conclusions}
\label{sec:conclusions}

In this letter, we derived a compact formula for the ratio $R_i=F_L^i/F_2^i$
of the heavy-flavour contributions to the proton structure functions $F_2$ and
$F_L$ valid through NLO at small values of Bjorken's $x$ variable.
We demonstrated the usefulness of this formula by extracting $F_2^c$ and
$F_2^b$ from the doubly differential cross section of DIS recently measured by
the H1 Collaboration \cite{Aktas:2004az,Aktas:2005iw} at HERA.
Our results agree with those extracted in
Refs.~\cite{Aktas:2004az,Aktas:2005iw} well within errors.
In the $Q^2$ range probed by the H1 data, our NLO predictions agree very well
with the LO ones and are rather stable under scale variations.
Since we worked in the fixed-flavour-number scheme, our results are bound to
break down for $Q^2\gg4m_i^2$, which manifests itself by appreciable QCD
correction factors and scale dependences.
As is well known, this problem is conveniently solved by adopting the
variable-flavour-number scheme, which we leave for future work.
Our approach also simply explains the feeble dependence of $R_i$ on $x$ and the
details of the PDFs in the small-$x$ regime.

\section*{Acknowledgements}

We are grateful to Sergei Chekanov, Vladimir Chekelian, Achim Geiser, Leonid
Gladilin, and Zakaria Merebashvili for useful discussions.
A.Yu.I. is grateful to the Scuola Internazionale Superiore di Studi Avanzati
(SISSA), where most of his work has been done.
A.V.K. was supported in part by the Alexander von Humboldt Foundation and
the Heisenberg-Landau Programme.
This work was supported in part by BMBF Grant No.\ 05 HT4GUA/4, HGF Grant
No.\ NG--VH--008, DFG Grant No.\ KN~365/7--1, and RFBR Grant No.\
 07-02-01046-a.

\end{document}